\documentclass[graybox]{svmult}

\usepackage{type1cm}        
\usepackage{makeidx}         
\usepackage{graphicx}        
\usepackage{multicol}        
\usepackage[bottom]{footmisc}

\usepackage{newtxtext}       %
\usepackage[varvw]{newtxmath}       

\makeindex             

\begin{document}

\title*{Hydrodynamics on (mini)superspace or a non-linear extension of quantum cosmology}
\subtitle{An effective timeless framework for cosmology from quantum gravity}
\author{Daniele Oriti\orcidID{0000-0002-2004-7063}}
\institute{Daniele Oriti \at Munich Center for Quantum Science and Technology 
and
Munich Center for Mathematical Philosophy,
Ludwig-Maximilians-University,
Schellingstrasse 4, 
D-80799 Munich, Germany, EU, \email{daniele.oriti@physik.lmu.de}}
%
%
\maketitle

\abstract*{We outline the content and theoretical support for the proposal of "hydrodynamics on (mini)superspace" (or a non-linear extension of quantum cosmology) as an effective framework for quantum gravity in a cosmological context. The basis for the proposal is a general correspondence between hydrodynamics and cosmology, and a picture of the universe as a quantum gravity condensate. The support comes from several directions. One is from mathematical physics: the hydrodynamics of quantum fluids can be mapped to relativistic cosmological dynamics, and both share the same conformal symmetries, which can be unraveled via geometric methods in superspace. 
The second is that the proposed framework is realized in quantum gravity formalisms like group field theory, in which an emergent cosmological dynamics can be extracted from the hydrodynamics of fundamental quantum simplices in a condensate phase. The same proposal can be motivated from the idea of 3rd quantization of gravity and from recent work on quantum cosmology, as an effective way of incorporating topology changing processes or cosmological inhomogeneities, respectively. 
A key conceptual ingredient is the relational understanding of space and time, which makes superspace the natural arena for gravitational dynamics, as opposed to the "spacetime" manifold, together with the general idea of emergent spacetime. 
The proposal and the results supporting it suggest an exciting dialogue between quantum gravity, the theory of quantum fluids and cosmology, as well as a new direction for analogue gravity simulations in the lab.
}


\section{Introduction}
\label{sec:1}
In this contribution we outline content, motivation and support for a new general framework for cosmology, understood as an effective description of an underlying theory of quantum spacetime and gravity, obtained upon suitable coarse graining of its constitutive structures.

\

A first set of motivations come from modern cosmology. Cosmology has produced a basic working account (the $\Lambda$CDM model of the cosmic history to an amazing level of detail, several theoretical models for its completion and improvements, and a large array of observations to test them.  A number of outstanding challenges remain, though. They include the problem of the cosmological singularity at the beginning (?) of our cosmic history, the choice of initial state of the universe, which affects the later evolution, the problem of transplanckian modes of cosmological perturbations in the early universe, and the dark energy (or cosmological constant) puzzle of the late (present) cosmological acceleration. Proposals of theoretical models solving one or more of these outstanding issues abound, but, in various ways, all need support or completion from a more fundamental theory of quantum gravity.

\

Let us give some examples. Models of the early universe are expected to provide an answer to two main questions: why do we have an approximately homogeneous, isotropic and flat universe? why an approximately scale invariant power spectrum for the cosmological perturbations? Answering these questions often ends up relying on assumptions about what happens at or close to the big bang singularity and more generally about quantum gravity physics \cite{cosmoQG} \cite{brandenberger}. In inflation, one has to clarify what produces the very accelerated expansion defining it (or the origin and nature of the inflaton, if such a new field is postulated); the physics of transplanckian modes may be made relevant by the very inflationary expansion, and is not fully under control by semiclassical field theory; also, if the inflationary phase is long enough, the relevant energy scale becomes dangerously close to the Planck scale where quantum gravity effects are expected to be dominant (and again not under control in the semiclassical framework); moreover, the fact that the inflationary universe still contains a singularity suggests its incompleteness and thus the possibility that whatever quantum gravity mechanisms resolves the singularity will affect the initial state of the universe as postulated in inflationary scenario or the features of the effective field theories used in it. In bouncing cosmologies, the need for the new physics of some quantum gravity theory to corroborate or modify the assumptions made about the bouncing phase is obvious, and the same can be said about the transition between the pre-big bang static phase and the post-big bang expanding phase in emergent universe scenarios (e.g. in string gas cosmology).  In the late universe, the accelerated expansion can in principle be accounted for, at least in terms of fitting observations, by a cosmological constant, thus simply a new parameter in a classical gravitational theory (unless cosmological tensions force us to adopt a dynamical vacuum energy picture even for purely data fitting purposes); however, its value depends on quantum fluctuations of gravitational and matter fields, and on their interaction, thus requiring a more fundamental theory of quantum gravity for achieving proper explanatory power \cite{CosmoConstant}. If dark energy is instead the result of a new type of field, then again a more fundamental theory is called for to ground it beyond the phenomenological level. If instead one opts for an account based on some modified gravity theory, this may be taken to represent a new fundamental theory of gravity, if it can be promoted to the quantum domain, or an effective classical-looking description of what is truly given by new quantum gravity effects at large scales \cite{DarkEnergy}. In turn this would imply that quantum gravity is not confined, as the effective field theory intuition would say, to the high energies/small distances domain. The latter hypothesis would be in line with emergent spacetime scenarios \cite{EmergentSpacetime}.

\

A second set of motivations comes from quantum gravity. There are many tentative quantum gravity formalisms, with different (and perspective-dependent) degrees of success and development. All of them have produced many mathematical results, conceptual insights and stimulating suggestions. They also offer many tentative indications of possible (and testable) phenomenology, including in cosmology. At the same time, they all face difficult outstanding challenges. Some of these challenges have to do with identifying and controlling the fundamental degrees of freedom of the theory and their quantum dynamics. Others have to do with controlling (mathematically and physically) the continuum, collective or macroscopic approximation of the theory and/or the reduction to effective spacetime physics in terms of quantum fields on (dynamical) geometric backgrounds. The second challenge includes the issue of connecting to cosmological dynamics starting from the fundamental quantum gravity formalism. A common strategy is to develop toy models that incorporate ingredients from the full theory into simplified contexts tailored to cosmological physics. Quantum cosmology is a well explored example. These toy models are very useful for guessing how the full theory could deal with cosmological issues, and can even produce viable predictions of new effects. However, lacking a clear embedding in or derivation from the full theory, their insights and their predictions are only partially compelling. Connecting cosmology to quantum gravity more fully, on the other hand, requires the development of proper approximation schemes within the full theory, which are likely to involve some form of coarse graining of quantum gravity degrees of freedom. 

\

Let us give an example of quantum gravity structures, hopefully clarifying why this should be the case. In a number of quantum gravity approaches, notably canonical loop quantum gravity, spin foam models, lattice quantum gravity (in connection variables) and tensorial group field theories, the fundamental quantum states are encoded in purely combinatorial and algebraic data: spin networks, i.e. graphs labelled by representations of Lie groups; and quantum histories of such states are expressed in the same language as spin foams, i.e. 2-complexes labelled by the same data. These combinatorial-algebraic structures can be obtained from the quantization of piecewise-flat or piecewise-degenerate geometries. In which case the algebraic data admit a discrete geometric interpretation, and reproduce such geometries for a subclass of \lq semi-classical\rq configurations and in specific approximations. These new \lq pre-geometric\rq structures are then subject to a quantum dynamics, which will then also be expressed without the usual spacetime language.  

This is the first general point to note: in the fundamental description of the universe and of spacetime, there may be no \lq\lq spacetime\rq\rq manifold, no spatial or temporal directions, and no fields as we usually take as basic ingredients of our physical models. It is in this sense that space and time may be emergent, in quantum gravity, i.e. they may be useful and physically salient notions with associated mathematical counterparts in some approximation only \cite{SpacetimeEmergence}. The second general point is that, given these fundamental structures, it is even more important to distinguish two conceptually, physically and mathematically distinct types of approximations one can perform. We can call them, for simplicity, the classical and the continuum approximations, even though both will take very different concrete forms in different formalisms (especially the second may not be directly expressed as a continuum limit in the usual sense of lattice theories). The first is the one taking from the quantum description of a given set of entities (particles, fields, etc) to a classical description of {\it the same entities}, restricting attention to states and dynamical regimes in which quantum effects are negligible (this is usually associated to \lq taking $\hbar$ to zero\rq). The second corresponds to restricting attention to a collective (dynamical) regime of (many of) the original quantum entities, loosing track of their individual properties and trading them for the properties of new averaged, collective entities. This second type of approximation may entail a form of classical limit with respect to the {\it new} collective entity, but depend crucially on the quantum nature of the original ones. Examples include moving from the physics of photons to that of the EM field, or from the theory of atoms to that of the quantum liquid they form upon condensation. Many examples can be exhibited to show that the two approximations/limits, classical and continuum, are physically distinct and {\it do not commute}, in general. We should expect the same for the fundamental quantum gravity entities and the approximations needed to recover spacetime geometry and fields from them \cite{Bronstein,LevelsEmergence}. 

\

All the motivations coming from theoretical cosmology and those coming from quantum gravity, lead us to focus our attention on the problem of the emergence of cosmological dynamics from fundamental quantum gravity, as a special case of the more general problem of the emergence of continuum spacetime physics from it. We ask then what could be the effective framework that captures cosmological physics including some quantum gravity effects, obtained from the fundamental theory upon coarse graining of its microscopic degrees of freedom (which could be non-spatiotemporal in nature).

\

The proposal we illustrate in the following is that cosmology can be connected to, in fact embedded in, quantum gravity via hydrodynamics. Specifically, we will argue that the effective framework we are looking for is given by \lq\lq hydrodynamics on minisuperspace\rq\rq. Minisuperspace can be understood as the configuration space of possible continuum field values at a single (manifold) point, or equivalently the configuration space of homogeneous fields. 

We will outline our proposed framework, hydrodynamics on (mini)superspace, in section ~\ref{sec:2}, together with its conceptual basis. Then we will offer the current evidence in support of this proposal. 
We will start, in section ~\ref{sec:3} with some intriguing results in mathematical physics, showing that standard hydrodynamics of superfluids can be mapped to relativistic cosmology, symmetries included, giving one specific example, and then indicating other known cases, on the one hand, and, on the other hand, how it is less surprising if the hydrodynamic system is understood as defined on minisuperspace. In section ~ref{sec:4} we will then summarize recent results in one quantum gravity formalism, i.e. tensorial group field theory (TGFT), showing how cosmological dynamics emerges exactly from its hydrodynamics  sector, defined on minisuperspace. The tensorial group field theory formalism is itself a convergence of several quantum gravity formalism, suggesting that the role of hydrodynamics on minisuperspace as the relevant effective framework for quantum gravity could be more general. In further support, we will then outline two other ways in which the same framework can be derived, from a quantum gravity perspective. 
Finally, in section ~\ref{sec:5} we will point out a possible important upshot of our proposal. If correct, it may indicate a new avenue for analogue gravity simulations in condensed matter systems. Indeed, the possibility of deriving cosmological dynamics from hydrodynamics (upon proper identification of its domain with minisuperspace) may imply that, contrary to conventional wisdom, analogue gravity simulations do not need to be confined to gravitational kinematics, but gravitational dynamics could be reproduced too, symmetries included, in the hydrodynamic regime of quantum fluids, and possibly also in the lab.

\section{Proposed framework and its conceptual basis}
\label{sec:2}
Let us illustrate the general aspects of our proposal, which, we argue, can provide cosmology with a mathematical completion within quantum gravity, and novel explanatory and predictive power, while conversely providing quantum gravity with a solid effective description and testing ground.

\subsection{General idea}
\label{subsec:2}

This mathematical framework we propose is given by hydrodynamics on (mini)superspace: a non-linear dynamics for a complex function (equivalently, its modulus and phase, encoding "fluid density" and "velocity") on the configuration space of spacetime fields "at a spacetime manifold point" (equivalently, homogeneous field configurations, or "minisuperspace", in the quantum gravity terminology\footnote{Minisuperspace is mathematically under control. For a general discussion of the much more complicated \lq superspace\rq, i.e. the full configuration space of (canonical) GR, see \cite{superspace}.}).

In such framework, cosmological quantities (scale factors, energy densities, etc) are hydrodynamic averages, satisfying dynamical equations derived from the hydrodynamic ones. 

Formally, hydrodynamics on minisuperspace can be seen as non-linear extension of quantum cosmology \cite{QC}, i.e. the straightforward quantization of  the homogeneous sector of GR. As already mentioned, quantum cosmology is a well-explored simplified framework for studying possible QG effects in cosmology, in need for a proper embedding in a complete QG formalism. In our proposal such embedding results in a non-linear (and non-local) extension, together with a radical shift in perspective.

Consider the variables of quantum cosmology\footnote{We restrict consideration to the 4-dimensional case.}. These are the metric data for a homogeneous spatial geometry, i.e. the three scale factors $a_i$ along three independent directions\footnote{One could extend the dynamical variables to include lapse and shift functions, in a canonical decomposition of the 4-metric. However, these are not dynamical degrees of freedom.}, and their conjugate variables corresponding to extrinsic curvature information. In addition, one can have several (homogeneous) matter fields, e.g. scalar fields, and their conjugate momenta. Minisuperspace, in this case, is the space of possible values of the three scale factors and of the scalar fields, thus $\mathbb{R}^3 \times \mathbb{R}^N$, where $N$ is the number of scalar fields, and we consider negative values of the scale factors to encode the same geometry of the positive values but with opposite orientation. The hydrodynamic variables defined on such domain can then be combined in a complex function $\Psi : mathbb{R}^3 \times \mathbb{R}^N \longrightarrow \mathbb{C}$. Consider the case of single matter scalar field for simplicity. The dynamical equations of hydrodynamics on minisuperspace will then take the general form:

\begin{eqnarray}
\label{eq:Hydro}
&& \mathcal{K}\left( a_i, \partial_{a_i}, \phi, \partial_\phi \right) \Psi(a_i, \phi) \, +\, \mathcal{V}^{'}\left[ \Psi\right] \, = \\ && = \,  \mathcal{K}\left( a_i, \partial_{a_i}, \phi, \partial_\phi \right) \Psi(a_i, \phi) \, +\, \lambda_3 \int  \Psi(a_{i_1},\phi_1) V_3(a_{i_1},a_{i_2},\phi,\phi_1) \Psi(a_{i_2},\phi_2) \,+ ....\,=\,0 \nonumber
\end{eqnarray}
that is, they will be integro-differential equations for $\Psi$ (and its complex conjugate) that will be generically non-linear and non-local on minisuperspace, with the precise form of non-locality encoded in the interaction kernels $V_i$, where we assumed that the non-linear functional of $\Psi$ can be expanded as a polynomial. The standard quantum cosmology formalism would correspond to a linear equation for the same $\Psi$, with the operator $\mathcal{K}$ given by the Hamiltonian constraint operator of a canonically quantized homogeneous sector of GR (or other classical gravitational theory). Here we are not assuming that this is the case for $\mathcal{K}$.

The new framework should be seen as a general coarse grained description of fundamental quantum gravity, not tied to a single QG formalism, in cosmological context (close to homogeneous backgrounds), incorporating (some) QG physics and inhomogeneities. We believe that it may represent a universal framework for the interface between quantum gravity and cosmology (thus, spacetime physics), incorporating a first set of quantum gravity effects, and that it could be derived from a variety of viewpoints and quantum gravity formalisms. We are going to give a variety of evidence in support of this belief, in the following, and we will see also a number of concrete realizations.

\subsection{Conceptual basis: relationalism and emergence}
Before doing so, we want to emphasize the conceptual basis of our proposal, some key perspectives that motivate, partially justify and provide a context to it.

\

{\bf Relationalism about spacetime} - In formulating this proposal, we are embracing a relational perspective on the construction of observables in a classical and quantum gravitational context, and on spacetime physics more generally. 

The main lessons from the diffeomorphism invariance and background independence of classical General Relativistic physics \cite{giulini} (which we assume is to be maintained in the quantum domain) are that there exist no absolute notion of temporal or spatial direction or location or distance, and that the \lq\lq spacetime manifold\rq\rq on which we routinely formulate spacetime physics has only a global role (providing a global restriction on the set of allowed geometries), i.e. that its local structures (points, directions, paths, coordinate frames, etc) have only a practical use but no physical significance; what is physically significant are the values of dynamical fields (among which the metric field) and relations among them. 

No function of manifold points, including the metric components, or the Ricci curvature scalar, or the value of a matter scalar field at a point, is an observable, because not invariant under diffeomorphism transformations. Quantities that are global with respect to the manifold, i.e. averages of scalar functionals of fields over the whole manifold (e.g. the total 4-volume), on the other hand, are technically invariant, thus formally observables, but they are clearly not very useful for local spacetime physics.  The relational strategy, which has a long history in gravitational physics (see \cite{Tambornino, Hoehn} for reviews), seeks to identify space and time in relations among dynamical fields, necessarily including the metric field (to provide a notion of spacetime extension), with one set of them, e.g. suitable matter fields, used as clock and rods to parametrize (in fact, {\i define}) the evolution and localization of the others. 

A schematic and simple example of the application of such strategy, in a context of spatially homogeneous fields, starts from two quantities like the (spatially homogeneous) Ricci scalar $R(t)$, function (via the homogeneous metric) of some temporal coordinate $t$, and a matter scalar field $\phi(t)$ with the same dependence, proceeds by inverting (when possible) this dependence to obtain $t(\phi)$ and then uses it to define the diffeomorphism-invariant observable $R(\phi)$ which is interpreted as providing the time evolution of the Ricci scalar as a function of the time measured by the clock $\phi$. 

Ideally, spacetime physics should only be expressed in terms of such relational quantities between fields like metric, gauge fields, matter fields, for both time and space localization. 
This is, in general, not possible in practice. Despite its long tradition, and the fact that the relational perspective is widely accepted to be {\it correct} as a matter of principle, among relativists and in the quantum gravity community, manifold points, coordinates, trajectories on the manifold etc are routinely used in gravitational physics. There is nothing wrong with this, of course, if only because it clearly works well. There is also no contradiction with the relational perspective: these structures are \lq useful fictions\rq ~that play (well) the role of physical frames in the approximation in which their actual physical properties (energy contribution and backreaction on other fields, quantum properties, etc) are negligible.

One lesson from the above is that physics and dynamics take place, strictly speaking, on superspace, i.e. the space of field configurations, not on the \lq spacetime manifold\rq, which is only a (useful) auxiliary structure. We do not have, however, a formulation of General Relativity (or other gravitational theories) purely in such relational language, i.e. in terms of equations only involving field values and no manifold, save very special cases. Things become much simpler in a cosmological context, where one can adopt the assumption of (approximately) homogeneous fields. 

In the homogeneous and isotropic restriction, for example, spacetime dynamics the metric only depends on the scale factor $a(t)$ and a lapse function $N(t)$ (which can fixed freely), with the scale factor encoding the universe volume (up to a constant) as $V = V_0 a^3$ (in the appropriate gauge). If the universe contains also a single free massless scalar matter field $\phi(t)$, the GR action reduces to $S = \frac{3}{8\pi G} \int dt N\left( - \frac{a V_0 \dot{a}^2}{N^2} + \frac{V}{N}\frac{\dot{\phi}^2}{2 N}\right)$, invariant under the 1d diffeos corresponding to reparametrizations of the time coordinate $t$. This should disappear from the relevant physics, which is fully captured by the diffeo-invariant relational observable $V(\chi)$ (obtained via the same \lq inversion step\rq mentioned earlier) and by the relational evolution $\left( \frac{1}{3V}\frac{dV}{d\phi}\right)^2 = \frac{4\pi G}{3}$ involving purely relations on the configuration space (minisuperspace) $\mathbb{R}\times \mathbb{R}$ defined by all values of the pair $\{ a,\chi\}$. The \lq time manifold\rq ~$\mathbb{R}$ with coordinate $t$ has disappeared from the picture.

The above relational formulation can be generalised to inhomogeneities and cosmological perturbations \cite{kristina}, and beyond, so it is not confined to toy models.

To summarize: to identify spacetime with the manifold supporting fields and spacetime physics with physics on that manifold is at best an approximation of a relational formulation in terms of correlations between fields, some of which used as reference frames. 

This conclusion has many conceptual and physical implications, only partially explored. From the point of view of quantum gravity, one main implication is that we do not expect to find the manifold and its structures at the fundamental level, nor in its effective description, if not introduced ad hoc for convenience.  

\

{\bf Spacetime emergence}  - As already discussed, the fundamental formulation of a quantum theory of spacetime and gravity may not be the result of a straightforward quantization of some classical gravitational theory like GR, and it may involve structures other than quantized continuum fields. These may actually be (as in canonical LQG, spin foam model and group field theory) purely combinatorial and algebraic structures instead, or discrete counterparts of continuum fields, as it is the case also in simplicial quantum gravity. 

Continuum spacetime and fields, and the (relational) geometric observables constructed from them would be the result of their collective quantum dynamics, and be \lq emergent\rq in the same sense in which fluid dynamics and fluid variables are emergent from the point of view of atomic physics. 

Notice that this applies to {\it all} GR structures and dynamics; this includes flat spacetime, which would be itself a highly excited, collective state of the \lq\lq QG atoms\rq\rq (we are thus beyond the \lq emergent gravity\rq picture, where gravitation and curved geometries possibly emerge from flat spacetime physics).

Indeed, we embrace a perspective on quantum spacetime as a peculiar (background-independent) quantum many-body system. Within it, the extraction of spacetime, and cosmology in particular,  is similar to the typical problem in condensed matter theory, i.e. the extraction of macroscopic, effective physics from the atomic one. Coarse graining procedures acquire then a central role, with the GR dynamics to appear as the approximate description of the collective quantum dynamics of many (infinite, in the idealized limit) QG atoms.

What should cosmology be, from this perspective? Cosmology is expected to correspond to the \lq most coarse-grained\rq dynamics of the QG atoms; in other words, it should be the effective dynamics of special global observables of the full theory, the result of a full coarse-grained description that restricts attention to such global quantities only. From the point of view of a quantum many-body system, this is the basic idea of a hydrodynamic regime. Thus, we may expect that cosmological dynamics will correspond to, or be extracted from, the quantum gravity hydrodynamics (and maybe share several features with the hydrodynamic description of standard fluids).

While this intuition needs to be substantiated by explicit derivations and much more evidence in order to be compelling, the above remarks should at least clarify what is the conceptual context within which our proposal takes form.

\

Now we move on to provide the evidence for it.

\section{Classical hydrodynamics/cosmology correspondence and its symmetry basis}
\label{sec:3}

We will now show that standard hydrodynamics can be mapped to relativistic cosmology, symmetries included. We will give some details on a few examples only, but the correspondence is more general. In fact, already the known examples hint at a deeper level of analysis, that could (and should) substantiate and explain the correspondence on physical grounds. Here, we will confine to the mathematical level.

\subsection{Hydro/cosmology map: general idea and example}

The general idea behind the mathematical correspondence is the following. Given a certain hydrodynamic system expressed as a scalar field theory, we can identify and compute a certain set of hydrodynamic averages, or integral moments of the distribution over the (single-atom) configuration space given by the fluid field. These hydrodynamic averages satisfy dynamical equations that are equivalent to the original hydrodynamic equations. Upon a precise correspondence between the hydrodynamic averages and cosmological variables (combinations of scale factors and matter energy densities), the dynamical equations for them can be shown to match.

Let us illustrate this idea with the earliest example of such correspondence, found in \cite{lidsey}. Consider the (cubic) Gross-Pitaevskii equation for a 2d, cylindrically symmetric, weakly interacting BEC (assuming a large conserved particle number), in flat space, described by a mean field $\Psi(r,t)$, given by:

\begin{equation}
i \hbar \frac{\partial\Psi}{\partial t} = - \frac{\hbar^2}{2 m}\nabla^2\Psi + V(r,t) \Psi + g | \Psi|^2 \Psi
\end{equation}
where $V(r,t) = m \omega^2(t) r^2 / 2$ is the trapping potential, $g$ is the interaction coupling among the atoms of mass $m$. The problem of solving this equation can be translated into solving the dynamical equations satisfied by the integral moments (representing the total particle number, width of the wave packet, radial momentum and energy of the system, respectively):

\begin{eqnarray}
&& I_1(t) = \int d^2 x |\Psi|^2 \qquad I_2(t) = \int d^2 x r^2 |\Psi|^2 \\ && I_3(t) = i \int d^2x r \left[ \Psi \frac{\partial \Psi*}{\partial r} - \Psi* \frac{\partial \Psi}{\partial r}\right] \qquad I_4(t) = \frac{1}{2} \int d^2 x \left[ |\nabla \Psi |^2 + g |\Psi|^4\right]
\end{eqnarray}
with the equations being:
\begin{equation}
\frac{d I_1}{dt} = 0 \qquad \frac{d I_2}{dt} = I_3 \qquad \frac{d I_3}{dt} = - 2 \omega^2 I_2 + 4 I_4 \qquad \frac{d I_4}{dt} = - \frac{1}{2} \omega^2 I_3
\end{equation}
and a conserved quantity $Q\equiv 2 I_4 I_2 - \frac{I_3^2}{4}$. 

In turn, these equations are equivalent to the Ermakov-Pinney equation $\frac{d^2 X}{dt^2} + \omega^2 X = \frac{Q}{X^3}$ for the variable $X = I_2^{1/2}$.

The dynamics encoded in these equations for the hydrodynamic averages can be mapped to the cosmological dynamics encoded in the Friedmann equations for a relativistic (homogeneous and isotropic) universe with spatial curvature $k$ filled with a scalar field with arbitrary potential. Let's see this map.

The Friedmann equations, and corresponding conservation law, expressed in cosmic time, look like:

\begin{equation}
H^2 = \frac{1}{a^2}\left( \frac{d a}{d\tau}\right)^2 = \frac{2}{3} \rho - \frac{k}{a^2} \qquad  \frac{d \rho}{d\tau} + 3 H (\rho + p) = 0
\end{equation}
where the energy density and pressure of the scalar field have the expression $\rho = \frac{1}{2}\left( \frac{d\phi}{d\tau}\right)^2 + U(\phi)$ and $p = \frac{1}{2}\left( \frac{d\phi}{d\tau}\right)^2 - U(\phi)$. When written in \lq laboratory time\rq ~$d/d\tau = a d/dt$, the Friedmann equations can be combined to give the same Ermakov-Pinney equation we had obtained for the BEC hydrodynamics: $\frac{d^2 a}{dt^2} + \left( \frac{d\phi}{d t}\right)^2 a = \frac{k}{a^3}$.

This means that one has a map between dynamical equations corresponding to a map between dynamical variables given by:
\begin{eqnarray}
I_2 \leftrightarrow  a^2 \qquad H^2 \leftrightarrow \frac{1}{4}\frac{I_3^2}{I_2} \qquad I_3 \leftrightarrow 2\frac{d a}{d\tau} \nonumber \\ \left( \frac{d\phi}{d\tau} \right)^2 \leftrightarrow \omega^2 I_2 \qquad I_4 \leftrightarrow \frac{\rho}{3} \qquad p \leftrightarrow I_2 \omega^2 - 3 I_4 \qquad Q = k 
\end{eqnarray}
with the matching of conservation laws:

\begin{equation}
\frac{d I_4}{d t} = - \frac{1}{2} \omega^2 I_3 \leftrightarrow  \frac{d \rho}{d\tau} + 3 H (\rho + p) = 0 \quad .
\end{equation}

The existence of such map is intriguing enough, but in itself it may sound like a mathematical coincidence only. 

Still, it applies to a number of other cosmological models and hydrodynamic systems. Indeed, similar maps to different kinds of BEC hydrodynamics have been found \cite{DAmbroiseWilliams, Gumjudpai, Gumjudpai2} to apply for: anisotropic Bianchi I cosmologies, Friedmann universes with cosmological constant, isotropic universes filled with matter satisfying different equations of state, inflationary universes for different inflaton potentials, universes filled with phantom-like dark energy, etc. 

\subsection{Matching symmetries}
In fact, the correspondence between BEC hydrodynamics and cosmological dynamics extends beyond the equations of motion to an exact match of symmetries. 

On the BEC side \cite{Ghosh, Horvathy, Cariglia}, for the same kind of fluid discussed above, the symmetry we have is the 2d centrally extended Schroedinger algebra $sh(2) = \left( sl(2,\mathbb{R}) \times U(1)\right) \ltimes \left( \mathbb{R}^2 \times \mathbb{R}^2\right)$, with generators (hydrodynamic averages):

\begin{eqnarray}
&& P_i  =  \int \rho \partial_i \theta \quad (translations) \qquad B_i = \int (\rho x_i - t P_i) \quad (Galilean \; boosts) \nonumber \\ &&J = \int e^{ij} x_i P_j \quad (rotations)  \qquad  Q_+ = \int \left( \frac{1}{2} \rho \gamma^{ij}\partial_i\theta \partial_j\theta + |\rho|^4 \right) \quad (Hamiltonian) \\ &&Q_0 = \int \left( t Q_+ - \frac{1}{2} x P\right) \quad (dilatations) \qquad Q_- = 2 t Q_0 - t^2 Q_+ + \frac{1}{2} \int \rho x^i x_i \quad (conformal \; transf) \, , \nonumber
\end{eqnarray}

and the central extension is given by the (conserved) total particle number.

What about the cosmology side of the supposed correspondence? It may seem impossible that the above Schroedinger symmetry is present there too. The GR dynamics has a symmetry given by the 4d diffeomorphisms, which, reduced by homogeneity and isotropy, leaves only time-reparametrizations or 1d diffeos as the only remnant spacetime symmetry. Indeed, the symmetry we are looking for cannot and should not be a \lq spacetime symmetry\rq i.e. a group of transformations acting on the \lq spacetime manifold\rq inducing then transformations on the dynamical fields. These are redundancies, for a generally covariant theory, and the manifold itself should disappear from a more fundamental formulation of cosmological physics, as we argued above. We are instead searching for symmetries with respect to transformations acting directly on field configurations, i.e. superspace, mapping one set of field values to another, without passing via the \lq spacetime manifold\rq. Such symmetries should exist beyond the \lq spacetime\rq transformations, in particular beyond any diffeomorphism invariance. 

Assuming that such symmetries may exist, how do we identify them? Two complementary methods are particularly aligned with the perspective we are taking here.  Both methods involving \lq geometrizing\rq the mechanical system of interest, and reformulating it as a dynamical system on the configuration space of mechanical variables (e.g. fields). We refer to \cite{BenAchourLivineOriti} for the general theory and for the application to cosmology and gravitational dynamics. Here we only outline the basic idea and results.

One method involves reformulating a general mechanical system for variables $\chi^a$ as the motion of a point particle on superspace with action:

\begin{equation}
S[\chi^a, \dot{\chi}^a] = c \int d\tau \left( \frac{1}{2}g_{ab}(\chi) \dot{\chi}^a \dot{\chi}^b - V(\chi) \right)
\end{equation} 

where the time coordinate plays now the role of affine parameter along the superspace trajectory, $g_{ab}(\chi)$ is the supermetric with line element on superspace $ds^2 = c g_{ab}(\chi) d\chi^a d\chi^b$, and we have gauge fixed the reparametrization invariance of the affine parameter. Equivalently, we can see the same system as the free motion of a point particle on the conformal superspace, upon conformal rescaling of the supermetric, and redefinition of the affine parameter: 

\begin{equation}
S[\chi^a, \dot{\chi}^a] = c \int d\eta \left( \frac{1}{2}G_{ab}(\chi) \dot{\chi}^a \dot{\chi}^b - 1\right) \qquad G_{ab} = V(\chi) g_{ab} \quad .
\end{equation} 

One can show that classical solutions of the original mechanical system correspond to geodesics of the conformal supermetric $G_{ab}$, and symmetries of the mechanical systems correspond to conformal isometries of the supermetric. From the point of view of superspace alone, free mechanical systems will correspond to null geodesics. For general potential, one has instead geodesics for positive mass, as it can be seen by restoring the reparametrization invariance and computing the associated constraint equation.

A second method is an immediate extension of the first, allowing to treat even interacting mechanical systems as free motion and to study efficiently also \lq time-dependent\rq symmetries, i.e. transformations that, while defined on superspace, depend in their definition on the \lq time coordinate\rq. It relies on the Eisenhart-Duval lift, an extended superspace augmented by two null directions, one of which corresponding to the original time coordinate, with line element $ds_{ED}^2 = 2 du dw - 2 V(\chi) du^2 + g_{ab}(\chi) d\chi^a d\chi^b$. Given the reliance on the time coordinate, this method does not account automatically for reparametrization invariance, which has to be imposed separately. The new point particle dynamics is then given by the action:

\begin{equation}
S[\chi^a, \dot{\chi}^a, u, w] =  \int d\lambda \left(\dot{u} \dot{w} - V(\chi) \dot{u}^2 +  \frac{1}{2}g_{ab}(\chi) \dot{\chi}^a \dot{\chi}^b \right) .
\end{equation} 

Now, classical solutions of the original mechanical system correspond to null geodesics of the ED-extended supermetric, and its symmetries (now including time-dependent transformations) correspond to its conformal isometries.  

Given these two general methods, one can investigate the hidden symmetries of cosmological (and other gravitational) systems \cite{BenAchourLivineOriti, BenAchourLivine}. For the Friedmann dynamics of a homogeneous and isotropic universe filled with a free massless scalar field, one identifies symmetry transformations in field space corresponding to all the ones found in the BEC hydrodynamics, i.e. the counterpart of translations (now in field space/superspace), Galilean boosts, conformal transformations, dilatation and rotations, and forming again the 2d centrally extended Schroedinger algebra $sh(2) = \left( sl(2,\mathbb{R}) \times U(1)\right) \ltimes \left( \mathbb{R}^2 \times \mathbb{R}^2\right)$. 

Interestingly, the central extension is now given by $n = \frac{2 V_0}{L_p^3}$, where $V_0$ is the volume of the fiducial cell of the cosmological model (over which one restricts spatial integrations, and thus playing the role of iR cutoff) and $L_P$ is the Planck length, playing the role of UV cutoff; in other words, it is given by the total (conserved) number of Planck-sized cells constituting the (homogeneous) universe with Friedmann dynamics. 

The explicit form of the field transformations and conserved charges/generators can be found in \cite{BenAchourLivineOriti}. For example,  the combination of translations and Galilean boosts can be written as:

\begin{eqnarray}
&&z \longrightarrow \tilde{z}(\tilde{\tau}) = z(\tau) + \frac{\xi(\tau)}{2} e^{\pm L_p} \phi/2  \nonumber \\ &&\phi \longrightarrow \tilde{\phi}(\tilde{\tau}) = \phi(\tau) - \frac{\pm \xi(\tau)}{L_p z} e^{\pm L_p} \phi/2  
\end{eqnarray}

where $z = a^{3/2}$ and $\xi$ is an arbitrary function of $\tau$, and the translation generators are:

\begin{equation}
P_\pm = e^{\mp L_p \phi/2}\left[ \frac{p}{2} \pm \frac{\pi}{L_p z}\right]
\end{equation}

with $p = - 2 n L_p N^{-1} \dot{z}$ is the conjugate variable to $z$ for generic lapse $N$, and $\pi = - \frac{n}{2} L_p^3 N^{-1} z^2 \dot{\phi}$ is the conjugate momentum of $\phi$.

A crucial point is that there is thus a precise unique correspondence between the conserved charges of the cosmological dynamics and the conserved charges of the hydrodynamic system; if the associated symmetry algebra is large enough, this allows to reconstruct fully one dynamical system from the other, and viceversa (if not, then the correspondence helps the reconstruction of some essential elements of the two, still). For example, the reconstruction is possible for the Friedmann/cubic BEC case, and one can re-derive the Lidsey correspondence entirely from the perspective of the Eisenhart-Duval lift \cite{Marcin2}. 
\

Despite the matching of symmetries too, that we have shown for the simplest FRW case, one could still see the hydrodynamics/cosmology correspondence as a mathematical coincidence only. We argue that there is more to it. Here we simply point out that also the symmetry-matching analysis can be generalised to a number of other gravitational systems, including Friedmann cosmologies with various matter scalar field potentials \cite{BenAchourLivineOriti}, Bianchi I cosmology \cite{Marcin}, and Schwartzchild-(A)deSitter black holes \cite{BenAchourLivineOriti2}.

\

In the next section, we discuss how the correspondence can be motivated and, in fact, derived from a more fundamental quantum gravity starting point, from a variety of perspectives.  

\subsection{Making sense of the correspondence} \label{subsec:MakingSense}
Before moving to quantum gravity considerations, let us stress how our proposal of hydrodynamics on superspace makes this hydrodynamics/cosmology correspondence intuitively reasonable. 

The correspondence appears natural, at least at the mathematical level, if one understands the BEC mean field $\Psi$ to be defined on (mini)superspace, i.e. as a function of the cosmological variables $(a,\phi)$, rather than \lq spacetime coordinates\rq $(r,t)$. This is not immediate for the original Lidsey map, because it relies on the time coordinate also on the cosmological side of the map. However, one gains an immediate intuition for the cosmological significance of the various elements of the hydrodynamic system when this is defined on minisuperspace and the cosmological quantities are defined in relational terms.

In fact, based on the symmetries, one can put in correspondence a 2nd order (\lq\lq relativistic\rq\rq) version of the BEC hydrodynamics with a purely relational formulation of cosmological dynamics, with a free massless scalar field used as a relational clock and no role for the temporal coordinate, with the first being defined on superspace only, while the Lidsey's map is obtained when lifting the domain of the hydrodynamics to the Eisenhart-Duval lift, including the temporal coordinate as a variable \cite{Marcin2}. 
In such rewriting, Lidsey's averages look like quantum cosmology averages for a mean field defined on an extended minisuperspace including the chosen time coordinate as a variable, i.e. $\Psi(a,\phi; t)$ and then foliated with respect to it. For example:

\begin{eqnarray}
I_2(t) &=& \int d a d\phi \, a^2 |\Psi(a,\phi;t)|^2 = \langle a^2 \rangle_{t} \longleftrightarrow a^2(t) \nonumber \\
I_3(t) &=& \int da d\phi \, a \left[ \Psi(a,\phi;t) \partial_a \Psi^*(a,\phi;t) - \Psi^*(a,\phi;t) \partial_a \Psi(a,\phi;t) \right] = \nonumber \\ &=& \langle a \pi_a \rangle_t = \langle a H \rangle_t  \longleftrightarrow (a H)(t)
\end{eqnarray}

Instead, in a purely relational context, one has a mean field $\Psi(a,\phi)$ defined on minisuperspace and integral moments, i.e. hydrodynamic averages, would look like standard quantum cosmology averages giving relational cosmological observables, with respect to the foliation (of minisuperspace) defined by the choice of relational clock given, say, by the scalar field. For example,  
\begin{eqnarray}
I_2(\phi) &=& \int d a \, a^2 |\Psi(a,\phi)|^2 = \langle a^2 \rangle_{\phi} \longleftrightarrow a^2(\phi) \nonumber \\
I_3(\phi) &=& \int da \, a \left[ \Psi(a,\phi) \partial_a \Psi^*(a,\phi) - \Psi^*(a,\phi) \partial_a \Psi(a,\phi) \right] = \nonumber \\ &=& \langle a \pi_a \rangle_\phi = \langle a H \rangle_\phi  \longleftrightarrow (a H)(\phi)
\end{eqnarray}

The same construction can be given for all the symmetry generators and other relevant observables.
For more details, see \cite{Marcin2}.

\

Following the above intuition, the maps between hydrodynamics and cosmology, discussed in this section give further motivation to our proposal.

\section{Quantum gravity support}
\label{sec:4}
The correspondence between classical cosmology and hydrodynamics, outlined above, even if shown to be very general, would not in itself vindicate our claim that hydrodynamics on superspace can represent a general effective framework for cosmology, encoding some quantum gravity physics. In order to do so, it is necessary to show a direct, if approximate, connection between this effective framework and fundamental quantum gravity formalisms, and possibly a concrete derivation of the the first from the second. It would be even better to show that there are many routes for such derivation, and several quantum gravity formalisms leading to the same effective framework, thus some degree of universality. This would make the theoretical aspects of the framework more compelling, but also any eventual observational consequence in cosmological physics more robust. 
This is what we do in this section: we should how cosmological dynamics emerges from quantum gravity in its hydrodynamics sector, defined on superspace, from a variety of viewpoints (but with different degrees of detail and robustness).  

\subsection{GFT condensate cosmology}
The derivation is rather complete and detailed in one specific quantum gravity formalism, namely tensorial group field theories, specifically the more quantum geometric models \cite{TGFT1,TGFT2} (simply referred to as \lq group field theories\rq). We first introduce the main elements of the formalism, then review very briefly how hydrodynamics on superspace and then cosmology emerge from it, and why this suggests that similar derivations are possible in related quantum gravity approaches. For more details on the derivation as well as on the many results of TGFT cosmology, we refer to introductions and reviews of the vast literature \cite{TGFTcosmo1,TGFTcosmo2,TGFT cosmo3}. See also \cite{TGFTemergence} for an extended discussion of how TGFT cosmology provides a concrete example of a more general template for the emergence of spacetime in quantum gravity.

\

These quantum geometric models describe the quantum structure of spacetime in terms of quantized tetrahedra, with algebraic data encoding their discrete geometry. The initial Hilbert space of an individual tetrahedron is $\mathcal{H} = L^2\left( G^4; d\mu_{Haar}\right)$, on which one imposes appropriate \lq geometricity \rq restrictions (which can be imposed at the level of dynamics). The Lie group $G$ is the Lorentz group $SO(3,1)$ (or its double cover $SL(2,\mathbb{C})$) or its rotation subgroup $SU(2)$, $d\mu_{Haar}$ is the Haar measure. Quantum states of geometry/spacetime are then functions of group elements associated to the triangles of each tetrahedron. An equivalent representation is in terms of (unitary, irreducible) group representations associated to the same triangles; thus, one can represent the same tetrahedron as a spin network vertex, i.e. a vertex with four outgoing open links labeled by group representations.  

One can then define a Fock space for arbitrary numbers of tetrahedra: $\mathcal{F}\left( \mathcal{H}\right) = \bigoplus_{V=0}^{\infty}sym\left\{ \mathcal{H}^{(1)}\otimes\mathcal{H}^{(2)}\otimes \cdots\otimes\mathcal{H}^{(V)}\right\}$, and introduce field operators creating/annhiliating the quanta of this Fock space, i.e. quantum tetrahedra, moving to a field-theoretic language.
In this Fock space, quantum states associated to extended simplicial complexes formed by gluing tetrahedra across shared boundary triangles correspond to (maximally) entangled states across the quantum degrees of freedom associated to the same triangle in two glued tetrahedra. Such states correspond to the spin network states associated to closed graphs in canonical loop quantum gravity and spin foam models \cite{LQG,SF}.

So TGFT states can encode discrete (piecewise-flat) geometries, at least for special states and in a semiclassical approximation. This discrete geometric intuition guides model building and the analysis of quantum dynamics. However, the list of \lq geometric or spatiotemporal pathologies\rq ˜that generic quantum states for arbitrary collections of TGFT building blocks can possess indicates the gap with respect to the usual description of spacetime in terms of fields (including a metric field). This justifies referring to this level of description as non-geometric and not spatiotemporal, and TGFT cosmology as an example of spacetime emergence.  

A partition function for the TGFT \lq atoms\rq, i.e. the quantized tetrahedra, will in general take the following field-theoretic form:

\begin{equation}
Z = \int \mathcal{D}\varphi\mathcal{D}\varphi^*\, e^{- \, S_\lambda(\varphi,\varphi^*)} \qquad ,
\end{equation}
for an action $S_\lambda(\varphi,\varphi^*) = K + U + U*$ function of the field (and complex conjugate) and some coupling constant(s) weighting an interaction term $U$ given by a polynomial in the fields, in addition to the quadratic kinetic term $K$. The pairing of their (group) arguments is, in general, non-local (they are not simply identified at interactions).

This form of the partition function can be motivated, when not derived, from a variety of viewpoints (see \cite{TGFTemergence} and references therein). A key fact, providing a useful guideline, is that it can be seen as the generating function, in its perturbative expansion, for a sum over simplicial complexes (which are dual to the TGFT Feynman diagrams) each weighted by a discrete gravity path integral (and incorporating also a sum over discrete topologies). For quantum geometric models, the lattice path integral associated to each Feynman diagram of the TGFT is highly non-trivial and formulated in algebraic discrete geometric variables, and in fact can be expressed equivalently as a spin foam model. Thus such TGFTs encode the continuum limit of lattice gravity path integrals and spin foam models, as well as the quantum dynamics of spin networks (quantum states of geometry in canonical loop quantum gravity), so that one can infer the appropriate TGFT action and partition function from knowledge of the lattice gravity or spin foam model they generate (and viceversa). 

This basic guideline can be used also to define TGFT models for quantum gravity coupled to matter. The strategy is to define a TGFT field and action in such a way that, in perturbative expansion, one obtains Feynman amplitudes with the form of discrete path integrals for gravity coupled to scalar fields, on the lattice dual to the Feynman diagram. This requires an extension of the domain of the TGFT field to include extra components corresponding to the possible values of the matter field, e.g. to $\varphi\left( g_I , \chi\right)$ for a single scalar field $\chi$, and a suitable coupling of these new data with the quantum geometric ones and across TGFT fields in the action.

\

The simplest approximation of the full quantum effective action, thus of the full continuum limit of lattice gravity path integrals and spin foam models, is the TGFT mean field hydrodynamics. It corresponds to the saddle point evaluation of the full TGFT path integral, and to approximating the full quantum effective action with the classical TGFT action $\Gamma[\phi] \simeq S(\phi)$. From the point of view of the QG atoms, i.e. the TGFT quanta, this amounts to working with highly quantum states: coherent states $| \Psi \rangle$ whose expression in terms of TGFT Fock excitations is:

\begin{equation}
| \Psi \rangle = \exp (\widehat{\Psi}) | 0 \rangle \qquad \widehat{\Psi} = \int d[g] d\chi \Psi(g_I;\chi) \widehat{\varphi}^\dagger(g_I;\chi) \qquad ,
\end{equation}

where $| 0 \rangle$ is the Fock vacuum (no QG atoms at all), and the exponential operator can be expanded to give an infinite superposition of states with increasing number of tetrahedra or spin network vertices. 
This is the quantum gravity counterpart of the Gross-Pitaevskii approximation  in the hydrodynamics of quantum liquids.

\

The task is to obtain, from such TGFT hydrodynamics, an effective dynamics expressed in spatiotemporal terms, in the sense of quantum GR, i.e. an effective dynamics of geometric quantities as those constructed out of continuum fields including the metric. 
When focusing on the TGFT hydrodynamics, we are assuming: a) that the relevant phase of the QG system is a {\it condensate phase} of QG entities, guided by the intuition of the universe as a kind of quantum fluid; b) that the relevant physical information is captured by the condensate wavefunction, i.e. the mean field $\Psi$; c) that a Gaussian, weakly interacting regime is already good enough to unravel interesting spacetime/gravitational physics.

The guess is that TGFT hydrodynamics will correspond to cosmological dynamics, for the heuristic reasons anticipated in the previous general discussion. Another reason follows from noticing that the TGFT mean field has the same domain of a wavefunction of a single GFT \lq atom\rq, i.e. an individual 3-simplex, encoding its (usually spacelike) quantum geometry \lq at a point\rq , i.e. the type of data that would suffice to describe cosmological dynamics. 

The guess is supported by a general fact, which applies to any TGFT model in which the domain $\mathcal{D}$ of the mean field $\Psi(\mathcal{D})$ (and of the fundamental field $\varphi(\mathcal{D})$) is understood as the space of geometries of a single (spacelike) 3-simplex (or conjugate extrinsic geometry), plus additional matter data. It can be shown \cite{Gielen,Jercher} that such domain $\mathcal{D}$ is diffeomorphic to the space of metrics (or conjugate extrinsic curvatures) at a point in a 3d (spacelike) hypersurface, plus matter field values at the same point, which in turn is diffeomorphic to the minisuperspace of continuum homogeneous 3-geometries (or conjugate homogeneous extrinsic data), plus homogeneous matter fields.
Therefore, the TGFT condensate wavefunction $\Psi(\mathcal{D})$ can be understood as a wavefunction on minisuperspace, as in quantum cosmology. 

The other general result is that this wavefunction satisfies non-linear dynamical equations, not the linear ones of quantum cosmology (i.e.the Wheeler-DeWitt equation restricted to wavefunctions on minisuperpsace). These are the quantum equations of motion derived from the quantum effective action or, in our approximation, the classical equations of motion of the chosen TGFT model (the quantum gravity analogue of a Gross-Pitaevskii hydrodynamic equation for a quantum fluid):

\begin{equation}
\int [dg'] d\chi' \mathcal{K}\left( [g] , [g']; \chi, \chi' \right)\,\Psi(g',\chi')\, + \, \lambda \frac{\delta}{\delta\varphi^*} \mathcal{V}(\varphi , \varphi^*) _{|_{\varphi = \Psi}} = 0 \qquad , \label{HydroEqns}
\end{equation}

with analogous equation for the conjugate TGFT field.

These equations are in general also non-local on minisuperspace, due to the non-local nature of the TGFT interactions. We obtain, then, a non-linear and non-local extension of a quantum cosmological equation, for the condensate wavefunction, encoding an infinity of quantum gravity degrees of freedom in a coarse grained, collective manner. It is an explicit and quite general realization of our proposed framework of hydrodynamics on (mini)superspace, derived at an effective coarse-grained level from a full quantum gravity formalism.

\

The next step is then to start from the hydrodynamic equations and extract from them dynamical equations for suitable geometric observables with a clear cosmological meaning. Of course, different TGFT models will give hydrodynamic equations that differ in their detailed form, and thus different cosmological dynamics.  We should also expect, however, some degree of universality across different models, since we are working an effective coarse-grained level.
This is confirmed by the analysis of specific TGFT models. 

Here, we report a sketch of the derivation of cosmological dynamics, highlighting the main steps without model-dependent details, and then summarize some interesting results obtained in specific cases.

A major simplification occurs when restricting attention to isotropic configurations, thus to functions of a single group representation label, once expanded in irreps of the group, and on the scalar field values only: $\Psi^j(\chi)$, corresponding to the fact that one single metric degree of freedom, e.g. the universe volume (or the scale factor), is relevant. A second simplification often applied is to neglect the TGFT interactions, i.e. the non-linearities of the equations, since consistency with lattice path integral (and spin foam) guidelines requires these interactions to be very weak. However, the effect of the interactions has also been studied, with interesting results. 

A key step is then to turn either the hydrodynamic equations directly or the ensuing equations for observables into relational form, by using one of the dynamical variables in the domain of the mean field as a relational clock. Often this role is played by the scalar field $\chi$, and there are several strategies to take this step, in the TGFT cosmology literature. One is to deparametrize the TGFT model from the start, i.e. at the level of the TGFT action, which results in a different (clock-dependent) Fock space than the one we gave above, different (clock-dependent) TGFT operators and observables, and hydrodynamic equations which are already written as temporal evolution equations, different from the general form we gave above. Here, we follow another route, and introduce instead an approximate, state-dependent relational framework.

We consider condensate states that are \lq semiclassical enough\rq with respect to the chosen clock variable, i..e the scalar field:

\begin{equation} 
\label{CPS}
\Psi_\epsilon(j;\chi) \equiv \eta_\epsilon(j;\chi-\chi_0;\pi_0)\tilde{\Psi}(j;\chi)\qquad ,
\end{equation}
where $\eta$ is a function (e.g. Gaussian) peaked around the $\chi_0$ value of the clock variable $\chi$, with width given by $\epsilon \ll1$, and depending on a second parameter $\pi_0$ governing the fluctuations in the conjugate variable to $\chi$ (related to the momentum of the scalar field, whose fluctuations are small if $\pi_0^2\epsilon \gg 1$).  
 
 The TGFT hydrodynamics equations are then well-approximated by equations for $\tilde{\Psi}$ involving \lq time\rq derivatives with respect to $\chi_0$, where in fact any higher order derivative (in $\chi_0$) possibly present in the original kinetic term can be neglected with respect to the second (and first) order ones:
 
 \begin{equation} \label{EffEqn}
\tilde{\Psi}^{''}_j(\chi_0) \, +\, A_j \tilde{\Psi}^{'}_j(\chi_0) \,-\,B_j \,\tilde{\Psi}_j(\chi_0)\,+ \, \mathcal{V}[\tilde{\Psi}]=\,0 \qquad ,
\end{equation}

where the coefficients $A_j$ and $B_j$ and the interaction functional $\mathcal{V}$ are, in general, function of both the parameters of the given TGFT model (the specific choice of TGFT action) and of the state ~\ref{CPS}.
 
One then needs to compute geometric observables, and specifically relational observables whose \lq temporal\rq localization is defined with respect to the scalar field clock $\chi_0$.

 These are expectation values of fundamental TGFT operators (acting on the TGFT Fock space) with a clear geometric interpretation (relying on the discrete gravity picture behind the TGFT model), evaluated on the states ~\ref{CPS}, and thus well approximated by the value the condensate wavefunction takes at $\chi=\chi_0$. Relevant ones are the occupation number: 

\begin{equation}
N(\chi_0) \equiv \langle \widehat{N}\rangle_{\Psi;\chi_0,\pi_0} = \sum_j \rho_j^2(\chi_0) \qquad ,
\end{equation}

the universe volume (constructed from the 1st quantized volume operator for quantized tetrahedra, with eigenvalues $V_j$:
\begin{equation}
V(\chi_0) \equiv \langle \widehat{V}\rangle_{\Psi;\chi_0,\pi_0} = \sum_j V_j \rho_j^2(\chi_0) \qquad ,
\end{equation}

the clock (scalar field) value:

\begin{equation}
\frac{\langle \widehat{\chi}\rangle_{\Psi;\chi_0,\pi_0}}{N(\chi_0)} \simeq \chi_0 \qquad ,
\end{equation}

and the scalar field momentum
$
\langle \widehat{\Pi}\rangle_{\Psi;\chi_0,\pi_0}$.  

All these quantities, which will represent the cosmological variables of the resulting cosmological dynamics, are hydrodynamic averages computed for a hydrodynamics system defined on (mini)superspace. This is exactly in line (despite differences in the actual derivation) with the hydrodynamics/cosmology maps discussed in section ~\ref{sec:3}.

\

Using the hydrodynamic equation ~\ref{EffEqn}, we then obtain the equations governing the relational evolution of the universe volume as a function of the energy density (defined via the above expectation values) of the scalar field (and of any other matter field we could have included in the model:
\begin{equation} \label{QuantumFriedmann}
\left( \frac{V'}{3V}\right)^2 = F\left( V, \rho, \theta, A_j, B_j, \mathcal{V}, \pi_0, \epsilon\right)  \qquad \frac{V^{''}}{V}  = G\left( V, \rho, \theta, A_j, B_j, \mathcal{V}, \pi_0, \epsilon\right) \qquad ,
\end{equation}
where of course the functionals $F$ and $G$ are again model- and state-dependent. 

These are the {\it generalised (quantum-corrected) Friedmann equations in relational time} for a given TGFT quantum gravity model, for the emergent spacetime in the homogeneous case, captured by the volume only, in the presence of the chosen simple matter content). 

The whole procedure can be generalised to different matter content. In particular, it can be generalised to deal with inhomogeneous universes, by introducing suitable matter frames (e.g. four free massless scalar fields) to localize observables in both space and time, in line with the relational strategy \cite{Marchetti}.

\

Many results obtained in TGFT cosmology over the last 10 years include the recovering of the classical GR dynamics for large volumes, the effective dynamics of cosmological perturbations in standard field theory language, the resolution of the cosmological singularity via a very robust bouncing mechanism, several mechanisms for producing an inflationary phase in the early universe or a (phantom-like) dark energy in the late universe, both of pure quantum gravity origin (no ad hoc matter fields), anisotropies, and more. We do not discuss them.

Our summary only intended to show how the framework of hydrodynamics on superspace, as an effective way of encoding cosmological dynamics, can in fact be derived from first principles in a candidate fundamental quantum gravity formalism. The viability, as well as the potential embedding within full quantum gravity, of the proposed framework should be now clear.

\

The question that remains is to what extent hydrodynamics on superspace can be seen as a universal description of a cosmological regime, shared across many different quantum gravity approaches.

Now we want to address at least partially this question, with two more examples of quantum gravity-related settings which also lead to the same framework. 

Before doing so, we stress that, in fact, already the derivation of hydrodynamics on superspace (and cosmology from it) from TGFTs suggests that it may represent a universal effective framework shared across different quantum gravity approaches. The reason is that the TGFT framework itself is a crossroad of different quantum gravity formalisms. TGFT models can be seen as an alternative definition of simplicial quantum gravity path integrals (involving a sum over triangulations) and, for quantum geometric models, as a completion of spin foam models for quantum gravity and a 2nd quantized reformulation of canonical loop quantum gravity. 

While the TGFT language may be particularly convenient for obtaining an effective coarse grained formulation of the quantum dynamics of spin networks, or of lattice gravity path integrals, it can also be taken to be simply an indication that such coarse grained formulation, even when achieved by other methods, will also give hydrodynamics on superspace as a result\footnote{Indeed, let us mention that effective continuum equations of this type, and in fact a formalism very similar to a simplified version of TGFTs, have been obtained also in the context of 2d simplicial quantum gravity formulated via causal dynamical triangulations, as the effective continuum limit of the lattice path integral extended to a sum over topologies, in \cite{Ambjorn}}.

\subsection{3rd quantization and effective topology change}
Let us now look at a different (but related) perspective on quantum gravity, not tied to discrete structures. Motivated mainly by difficulties in defining a positive definite inner product in canonical quantum gravity (due to the mixed signature of the supermetric) and by the will to incorporate topology change, the 3rd quantization idea was proposed over thirty years ago \cite{3rd}. The idea has surfaced back recently, due to the renewed interested in wormhole production in quantum gravity \cite{wormhole}. It has never gained too much traction, mostly because making mathematical sense of it is even more challenging than for the gravitational path integral (or canonical quantization) itself. It has a great conceptual appeal, however, and one can also see TGFTs (and the earlier matrix and tensor models) as a concrete realization of it in a discrete gravity context \cite{DiscreteContinuum3rd}.

The basic idea is to define a \lq quantum field theory of universes\rq by promoting the canonical gravity wavefunction to a field $\Phi[^3 g]$ living (like the canonical wavefunction) on superspace, i.e. the space of 3-geometries for given spatial topology (we assume that of a 3-sphere).  

The dynamics of this superspace field (assumed real) is encoded in the action:

\begin{equation} \label{3rd}
S = -\frac{1}{2} \int \mathcal{D} ^3 g\, \Phi \square \Phi \, +\, V[\Phi] =  -\frac{1}{2} \int \mathcal{D} ^3 g\, \Phi \square \Phi \, \, +\,\lambda_3 V_3[\Phi] + .... 
\end{equation}
where the kinetic kernel is given by the Wheeler-DeWitt Hamiltonian constraint of canonical gravity, and the interaction terms encode topology changing processes. For example, $V_3$ governs the splitting of a universe into two with the same topology, i.e. a \lq trousers\rq process or its inverse (merging of two universes into one), with a form like:
\begin{equation} 
V_3 = \int \mathcal{D} ^3 g_1 \mathcal{D} ^3 g_2 \mathcal{D} ^3 g_3\, \Phi[^3 g_1] \Phi[^3 g_2] \Phi[^3 g_3] \delta ( ^3 g_2 , ^3 g_1^-) \delta ( ^3 g_3 , ^3 g_1^+)
\end{equation}
where the kernel gives matching conditions at the topological junctions across universes.

The classical equations of motion of this theory, $\frac{\delta S}{\delta\Phi} = \box \Phi = V^{'}[\Phi] = 0$ would then be by definition dynamical equations for a wavefunction on superspace and a non-linear and in general non-local extension of the Wheeler-DeWitt equation of canonical quantum gravity. They will represent the simplest (mean field) approximation of the full quantum equations of motion of the theory, obtained by variation of the quantum effective action $\frac{\delta \Gamma[\Phi_B]}{\delta\Phi_B} = 0$, which will be the fully quantum-corrected non-linear (and non-local) extension of the Wheeler-DeWitt equation, including the effects of topology change.

The quantum field theory defined by the action ~\ref{3rd}  would have a partition function which, in a perturbative expansion, would give a sum over Feynman \lq diagrams\rq dual to manifolds of higher an higher genus, obtained connecting a propagator corresponding to a cylindrical, globally hyperbolic topology (with boundary 3-manifolds being 3-spheres, with interaction vertices corresponding to trousers topologies (assuming we only include the interaction $V_3$ of both orientations:
\begin{equation}
Z_{\lambda_3} = \int \mathcal{D}\Phi[^3 g] \, e^{- S[\Phi]} = \sum_{\mathcal{M}} \mathcal{A}[\mathcal{M}] \qquad .
\end{equation}

Each perturbative process would be weighted by a quantum gravity path integral on the given topology and involving the classical gravity action corresponding to the Wheeler-DeWitt operator $\square$, e.g. the Einstein-Hilbert action of GR:
\begin{equation}
\mathcal{A}[\mathcal{M}] = \int_{\{ g |\mathcal{M}\}} \mathcal{D} g\, e^{i S^{EH}_{\mathcal{M}}(g)}
\end{equation}

\

Obviously, the theory so formulated in entirely formal and faces enormous mathematical difficulties. Already the functional measure of the gravitational path integral, entering the perturbative amplitudes as well as the action, is a major obstacle (properly defining it would amount to solving a major difficulty of canonical and covariant quantum gravity). It is no surprise, then, that achievements in this framework have been confined to conceptual insights, formal computations suggesting possible new physics, and semiclassical or otherwise approximate results. We notice once more, however, that matrix models for 2d gravity, and tensor models and TGFTs in higher dimensions, can be seen as a more rigorous formulation in a discrete context, but in the very same spirit, and allowed much more progress \cite{DiscreteContinuum3rd}.

Among the simplifications that one can adopt to make the 3rd quantization framework more rigorous, there is also the minisuperspace reduction to homogeneous geometries (and matter fields) \cite{3rdMini}. This reduces the framework to that of an ordinary scalar field theory on minisuperspace. In the case of isotropic universes governed by GR at the classical level and filled with a single matter scalar field, one has a 2d minisuperspace with coordinates $\{ a, \phi\}$, and an action:
\begin{eqnarray}
S &=& \frac{1}{2} \int d a d\phi\left[ \Phi \,\partial_a^2 \Phi - \frac{1}{a^2}\Phi \,\partial_\phi^2 \Phi + (a^2 - a^4 - a^4 U(\phi) \Phi^2\right]  \\ &+& \frac{\lambda_3}{2} \int da da^{'} da^{''} d\phi d\phi^{'} d\phi^{''} \Phi(a,\phi) \Phi(a^{'},\phi^{'}) \Phi(a^{''},\phi^{''}) \mathcal{V}\left( a, \phi; a^{'}, \phi^{'}; a^{''}, \phi^{''}\right) \nonumber
\end{eqnarray}
 
This is exactly the sort of hydrodynamics on minisuperspace we have discussed in the previous sections, and the object of our proposal. Indeed, the corresponding equations of motion give a non-linear (and non-local) extension of the dynamics of quantum cosmology. Here, the non-linearities would encode, at an effective level, the physics of topology changing processes, in a language that only refers to a single, homogeneous universe.

\

It should be clear that the 3rd quantization idea does not depend on the specific formulation of classical or quantum gravity/cosmology one has at hand. Indeed, the minisuperspace version of 3rd quantization has been proposed and analysed \cite{3rdLQC} having as starting point loop quantum cosmology, based on connection and tetrad variables. In the variables used in the loop quantum cosmology framework, in particular $\nu$ labelling the eigenvalues of the volume operator, one has an action

\begin{eqnarray}
S_{{\rm i}}[\Phi]&=&\frac{1}{2}\sum_{\nu}\int d\phi\;\Psi(\nu,\phi)\hat{\mathcal{K}}\Phi(\nu,\phi)+\sum_{j=2}^n\frac{\lambda_j}{j!}\times\label{intact}
\\&&\sum_{\nu_1\ldots\nu_j}\int d\phi_1\ldots d\phi_j\; f_j(\nu_i,\phi_i) \prod_{k=1}^j\Phi(\nu_k,\phi_k)\,,\nonumber
\end{eqnarray}

for generic interaction kernels and a kinetic operator being the Hamiltonian constraint operator in (isotropic) loop quantum cosmology, given by a difference operator

\begin{eqnarray}
\hat{\mathcal{K}}\,\Phi(\nu,\phi) := -B(\nu)\left(\Theta+\partial_{\phi}^2\right)\Phi(\nu,\phi)= 0\,,
\\
-B(\nu)\Theta\psi(\nu,\phi) &:= &A(\nu)\psi(\nu+\nu_0,\phi)+C(\nu)\psi(\nu,\phi)\nonumber
\\& & +D(\nu)\psi(\nu-\nu_0,\phi)\,.
\end{eqnarray}
Here $A,B,C$, and $D$ are functions depending on the details of the quantization scheme and the choice of lapse function), and $\nu_0$ is an elementary length unit, usually defined by the square root of the area gap (proportional to the Planck length).  

The non-linearities have again the interpretation of encoding topology changing processes, in a 3rd quantization perspective. Different specific choices and conditions one may impose on such interactions are discussed in \cite{3rdLQC}.

In fact, another interpretation is possible for the same interactions, following which one would make different choices for the interaction kernels, with the same general form of the dynamics. 

Instead of a field theory of interacting homogeneous universes, one could interpret it as a field theory of interacting homogeneous {\it patches} of a single inhomogeneous universe. The interactions would then govern processes in which such patches merge or split, giving an effective way of dealing with cosmological inhomoegeneities. This option is also discussed in some detail in \cite{3rdLQC}.

It is also the perspective taken in following another route that leads to hydrodynamics on minisuperspace and non-linear quantum cosmology, which we now discuss.

\subsection{Separate universe and effective inhomogeneities}
In \cite{Martin}, the starting point is the separate-universe approach to cosmological perturbations. In this approach \cite{separate}, the universe is subdivided into homogeneous regions with different geometry. Whether this is a good or too coarse description of an inhomogeneous universe depends on its precise geometry. One expects that for strong curvature and small wavelengths of the cosmological perturbations, one has to use homogeneous patches with very small volume, and thus very many of them (for given total 3-volume of the universe), in order to have a good approximation at both kinematical and dynamical level. For low curvature and long wavelengths of cosmological perturbations, large homogeneous patches would suffice. The relative differences in cosmological variables, e.g. 3-volume or scale factors, would then measure the degree of inhomogeneity across patches.

In the limiting case of very strong curvature, e.g. close to the cosmological singularity, one would have a (infinite) number of extremely small homogeneous patches, one per spatial manifold point in the actual limit. Each patch would evolve according to the dynamics of homogeneous cosmology. The BKL conjecture would then state that the individual patch dynamics for all patches is actually enough to control the full universe dynamics, i.e. that spatial gradients (relations across patches) are in fact negligible. 

Let's take this starting point serious, but confine ourselves to the less extreme situation of many homogeneous patches of finite size, individually governed by a homogeneous dynamics but with non-negligible interactions among them. The analysis of \cite{Martin} aimed at extracting an effective coarse grained dynamical framework for such universe, following the example of BECs, and in the form of a hydrodynamics of homogeneous patches. We will see that this turns out to be a hydrodynamics on minisuperspace and a non-linear extension of quantum cosmology. We sketch here the main steps of the derivation, and refer to \cite{Martin} for more details.

\

We split the spatial manifold $\Sigma$ into several homogeneous and isotropic regions: $\Sigma = \bigcup^{N^{1/3}}_{i,j,k = 1} \mathcal{S}_{i,j,k}$, where the notation suggest a cubic lattice but we need to make no assumption about the precise topology of the decomposition. Because of homogeneity and isotropy, for each region it is enough to consider only the volume degree of freedom $V_{i,j,k}$, such that $V = \sum_{i,j,k} V_{i,j,k}$. For simplicity, we neglect any matter content of the universe, in the following.

The dynamics is encoded in the Hamiltonian constraint of GR, which is adapted (discretized) to the regions as:

\begin{eqnarray} \label{H}
H_{disc} = - 6 \pi G V \left( \Pi^2_{V} + \frac{1}{N}  \sum_{i,j,k} \Pi_{i,j,k}^2 + ...\right) \\ \Pi_V = - \frac{1}{12\pi G}\frac{\dot{V}}{V} \qquad \Pi_{i,j,k} = - \frac{1}{12\pi G} \dot{\left( \frac{N V_{i,j,k}}{V}\right)} \quad \nonumber .
\end{eqnarray} 

where we have neglected higher derivatives terms, thus assuming inhomogeneities to be small.

Inhomogeneities are indeed captured by differences in patch volumes $V_{(i,j,k) + b} - V_{(i,j,k) - b}$, where $b$ is a unit vector in the b-direction (in some coordinate system). The dynamics will include, in general, interaction terms  given by (complicated) polynomials of such differences.

 We can then quantize each patch as in (loop) quantum cosmology, e.g. in the volume representation, obtaining for each patch a wavefunction $\psi_{V_{i,j,k}}$.

The total state of the inhomogeneous universe could be then considered, treating the universe as a quantum many-body system with each patch playing the role of a fundamental atomic constituent. Notice the resonance with the TGFT framework. 

If inhomogeneities are small, thus correlations among patches are small, then a reasonable approximation for the total state of the universe is the product state:

\begin{equation}
\Psi\left( V_1,V_2, ....\right) = \frac{1}{N !} \psi_1(V_1) \psi_2(V_2) \cdots
\end{equation}
where we have assumed a relabelling invariance of the patches (i.s. a bosonic statistics). This can be seen as a counterpart of diffeomorphism invariance in a discrete setting, and it is a common element in canonical loop quantum gravity and TGFTs, as well as in simplicial quantum gravity.

Incidentally, this form of the total wavefunction for the universe has been suggested also in \cite{Nicolai}, in the context of a canonical quantization of (symmetry-reduced) supergravity based on symmetries, for the regime close to the cosmological singularity, inspired by the BKL conjecture.

In fact, if inhomogeneities are -very- small, then one could expect the quantum state of the universe to be close to an even simpler state, of the form:

\begin{equation} \label{condensateLQC}
\Psi\left( V_1,V_2, ....\right) = \frac{1}{N !} \psi(V_1) \psi(V_2) \cdots \qquad ,
\end{equation}
that is, a condensate state in which all atomic constituents, i.e. all homogeneous patches, have the same individual wavefunction. Notice that this is exactly the kind of quantum geometry states used to extract the mean field approximation in TGFT cosmology, that is the coherent states of the TGFT field, now restricted to the case of a fixed \lq particle number\rq, i.e. a fixed number of homogeneous patches (there, in TGFT, QG atoms).

Following these assumptions, we now need to evaluate the quantum dynamics encoded in the Hamiltonian constraint  ~\ref{H}. We can proceed as in the case of BECs. The simplest approximation is a quantum dynamics obtained from minimizing th expectation value of the Hamiltonian constraint evaluated in the full quantum  state ~\ref{condensateLQC}. This is akin to minimizing the energy functional in BECs. In such quantum state, the single-patch wavefunction $\psi$ determines the whole state as a collective variable, and the resulting dynamics will necessarily be some equation for it.

Proceeding in this way, the expectation value of the quadratic term in the Hamiltonian constraint gives a quadratic functional in $\psi$ with the homogeneous WdW operator (in loop quantum cosmology formulation) as its kernel.
The interaction (higher-order) terms give instead a higher-order functional in $\psi$. For example, considering the interactions between pairs of homogeneous patches only:

\begin{eqnarray}
W_{int}(V_1,V_2) = \alpha (V_1 - V_2)^2/V^2 \qquad &\longrightarrow& \nonumber \\ \longrightarrow \qquad \langle \hat{W}_{int} \rangle_{\Psi} &=& \frac{\alpha}{V^2} \int dV_1 dV_2 |\psi(V_1)|^2 |\psi(V_2)|^2 (V_1 - V_2)^2 \nonumber
\end{eqnarray}

It is clear, then, that the end result of this construction is going to be a non-linear (and non-local) equation for the collective variable given by the cosmological wavefunction $\psi$ defined on minisuperspace, with specific details depending on the initial Hamiltonian constraint operator and the various approximations ad choices made along the way. In fact, the resulting dynamics will be very close to the one presented in the previous section, coming from 3rd quantized loop quantum cosmology.

In the end, we obtain again, from a different perspective, an effective framework given by hydrodynamics on (mini)superspace, further supporting our proposal.

\section{A possible upshot: a new route for analogue gravity?}
\label{sec:5}

Before concluding, let us point out a further possible upshot of our proposal: a novel research direction for analogue gravity simulations in condensed matter systems. Specifically, it suggests that such simulations do not need to be confined to the kinematical sector of gravitational theories: the gravitational dynamics can be reproduced too, possibly beyond the classical regime. 

Analogue gravity models in condensed matter systems are by now an established research area, backed up by a large body of literature; see \cite{analogue} for a review. They are the subject of a variety of theoretical developments, exploring the details of the gravitational interpretation of condensed matter phenomena, and of promising experimental efforts, aiming at reproducing in the lab the phenomena corresponding to often extreme (semi-classical) gravitational effects which are difficult or even impossible to test directly by astrophysical or cosmological observations.  

They are based on the fact that an effective curved geometry emerges naturally and generically at the hydrodynamic level in condensed matter systems, specifically classical and quantum fluids (in fact, this fact extends also to a number of non-fluid systems). More precisely, excitations over stable ground states of these systems couple to this emergent geometry, and not to the (background) laboratory metric, and their effective description is given by (relativistic) QFTs on curved spacetimes, with the speed of sound playing the role of the speed of light in the standard spacetime context. Moreover, the correspondence is rather general also from a gravitational perspective, in the sense that the class of emergent curved geometries is broad enough to include a large number of interesting gravitational phenomena, in particular for semi-classical effects around black holes and in cosmological spacetimes of direct physical interest.

As mentioned, one reason for the attention devoted to this research area is that analogue gravitational scenarios, including exotic ones like Hawking radiation or cosmological particle production, can be reproduced (and measured) in concrete lab situations. A more theoretical reason of interest is that condensed matter systems allow investigate frontier scenarios (e.g. violation of Lorentz symmetry or other possible quantum gravity phenomenology), as well as more formal aspects of classical and quantum gravity (e.g. the role of the equivalence principle, quantum unitarity etc, of interest for fundamental physics), with (in principle) total control over both \lq effective curved spacetime QFT\rq ~and the \lq fundamental\rq atomic physics originating it. This is lacking in fundamental physics, where the underlying quantum gravity theory is not under control nor established.

However promising all of the above is (see again \cite{analogue} for a proper discussion of achievements, current research and open issues), one big issue limits the actual impact of results and prospects in the field, as currently understood. The gravitational dynamics is not reproduced in analogue gravity system. In fact, the (almost complete) consensus is that the gravitational dynamics just {\it cannot} be reproduced, at least as described by a generally covariant theory like GR.

The physical reasons for this consensus are strong. They derive from the structural core of the gravitational analogy in condensed matter: the dynamics of emergent geometry is governed by the non-relativistic hydrodynamic eqns of the system. These are structurally different from GR-like eqns, and the structural differences encode crucial physical differences. 
In particular:  hydrodynamics is background dependent, i.e. it depends on a fixed background spacetime metric, and thus not fully dynamical from the spacetime point of view; it identifies preferred frames (isometries of background metric); it has a special global symmetry, i.e. the Schroedinger symmetry (which can be extended to the relativistic case to include Lorentz boost, rather than Galilean ones) rather than the diffeomorphism symmetry of GR and other generally covariant gravitational theories.

All these differences may be obscured or negligible at a semi-classical, perturbative level (even gravitational perturbations can in principle be reproduced in analogue systems), but eventually show up both at the theoretical level and, one should expect, the experimental one. For example, they prevent a full account of the physics of the backreaction of (quantum) fields on spacetime itself.

This is a strong constraint on the potential development of analogue gravity simulations and of their theoretical analysis. It is also, for many, a strong reason to warrant only limited value to the many results obtained in this area, or to maintain a certain mistrust on their fundamental significance. 

This is not the place to attempt any detailed analysis, and at the moment we do not have enough results to properly undermine this consensus. However, it should be clear that, at the very least, our results from Quantum Gravity (TGFT) and mathematical physics, and the relational perspective on gravitational physics, suggest otherwise, and our proposal indicates a possible way forward. 

Consider the derivation of cosmological dynamics (including cosmological perturbations) from quantum gravity hydrodynamics; and consider the existing maps between cosmology and BEC hydrodynamics and their matching symmetries. These results, discussed above, and more that can be found in the literature, show that hydrodynamic equations can, in fact, be mapped to relativistic cosmological dynamics, i.e. what is argued to be impossible in the analogue gravity context. Similarly, the two dynamical contexts have in fact the same symmetries, once properly compared. 

The key point is to accept a shift in perspective.  The hydrodynamics should be understood as defined on (mini)superspace (the space of field values), and not on the \lq spacetime manifold\rq, and the relevant symmetries are field transformations on (mini)superspace, and not on the \lq spacetime manifold\rq. Then the reliance on a metric background with special isometries is not in contradiction with the general covariance and background independence (from the point of view of spacetime physics) of GR: (mini)superspace is a metric manifold with a fixed metric, the supermetric, which has non trivial isometries, those corresponding to the field symmetries of the gravitational dynamics. 

Thus, it may be possible to have a proper gravitational dynamics (for both background universe and perturbations) in quantum many-body systems, in a hydrodynamic approximation of the same, after all, provided one changes to a relational or "superspace-based" perspective on spacetime physics, rather than assigning a central role to the \lq spacetime manifold\rq, which is, in fact, a redundant mathematical structure. 

The suggestion for analogue gravity simulations would be, then, to try to reproduce in the lab not the \lq spacetime manifold\rq, its metric and the fields living on top of it, but (mini)superspace with its (super) metric, and the appropriate hydrodynamic system living on it. Having done so, cosmological observables and, more generally, gravitational dynamics should be reconstructed in a second step, from appropriate hydrodynamic averages, rather than identified directly with hydrodynamic perturbations. Specifically, one could focus on the examples of minisuperspace hydrodynamics discussed in the previous sections, that have been already shown to allow a reconstruction of gravitational dynamics, and on any other such models as suggested, e.g., by quantum gravity considerations.

We leave a more detailed elaboration of the above points to another contribution, and the proper implementation of the suggestion in an analogue gravity context to future work.

\section{Conclusions} 
We have put forward a proposal for an effective framework to study cosmological physics, incorporating quantum gravity features of our universe, given by hydrodynamics on minisuperspace (the space of field configurations \lq at a point\rq or the space of homogeneous fields). This hydrodynamics on minisuperspace amounts to a non-linear and non-local extension of quantum cosmology, but with a very different perspective. It is understood, in fact, as the result of some coarse graining of quantum gravity degrees of freedom, which may well be non-spatiotemporal and non-geometric, rather than of a straightforward quantization of a symmetry-reduced sector of a classical gravitational theory (thus, while there may be an underlying Hilbert space of \lq QG atoms\rq, there is no Hilbert space of homogeneous universes, as in quantum cosmology). 

\

We have outlined the conceptual basis of the proposal, built on a relational perspective on spacetime and on the idea of spacetime emergence, combined with a vision of the universe as a peculiar quantum many-body system. We then offered some evidence in support of the proposal. We started from a number of results in mathematical physics, unravelling a surprising correspondence between cosmological dynamics and standard hydrodynamics.Then, we summarized recent results in one quantum gravity formalism, i.e. tensorial group field theory (TGFT), showing how cosmological dynamics emerges from its hydrodynamics  sector, defined on minisuperspace. The tensorial group field theory formalism is itself a convergence of several quantum gravity formalisms, suggesting that the role of hydrodynamics on minisuperspace as the relevant effective framework could be more general. In further support of this hypothesis, we have outlined two other ways in which the same framework can be derived, in a quantum gravity context, one following the idea of 3rd quantization of gravity, as an effective way of accounting for topology change, and another based on techniques from theoretical cosmology as well as (loop) quantum cosmology, as an effective way to deal with cosmological inhomogenities.  

\

While the research programme of TGFT cosmology is rather established and already progressing fast, with more results yet to come but already within the radar, much more work is needed to establish the proposed framework as a universal one across quantum gravity approaches. Such universality, if shown convincingly, will add to the theoretical interest in the proposed framework, as well as to the robustness of its predictions of new cosmological physics. Beyond the theoretical developments, it is the direction of physical cosmology, in fact, that should be pursued intensively, in order to prove the usefulness of the proposed framework in producing testable predictions of new cosmological phenomena, of possible quantum gravity imprints in them, or novel explanations of existing cosmological puzzles. The implementation of the suggested new route for analogue gravity simulations reproducing also the gravitational dynamics should be, of course, another objective to pursue.

Along any of the above research directions, we expect an exciting dialogue between quantum gravity, the theory of quantum fluids and cosmology.

\begin{acknowledgement}
We gratefully acknowledge useful discussions and collaborations with J. Ben Achour, E. Colafranceschi, R. Dekhil, S. Gielen, A. Jercher, S. Langenscheidt, E. Livine, L. Marchetti,  X. Pang, A. Pithis, M. Sakellariadou, M. Stankiewicz, J. Thuerigen, Y. Wang, E. Wilson-Ewing. This work has been supported by the Deutsche Forschung Gemeinschaft (DFG) and by the Templeton Foundation. 
\end{acknowledgement}
%

%
%
%

\end{document}